\documentclass[journal]{IEEEtran}
\usepackage{cite}
\usepackage{graphicx}
\usepackage{amsmath}
\usepackage{algorithmic}
\usepackage{authblk}
\usepackage[hyphens]{url}
\usepackage{float}

\begin{document}
\title{PaddleSat Optical Charging Station in Space}
\author{
Anish Nair, Austin O'Connell, Shalomi Arulpragasam, Noah Kim%
\thanks{Emails: \texttt{anair323@gatech.edu, coconnell7@gatech.edu, sjnimesh@gatech.edu, nkim336@gatech.edu}}
}

\maketitle

\begin{abstract}
This work investigates the feasibility and design trade-offs for a companion spacecraft, or PaddleSat, to charge a host spacecraft by wirelessly transmitting power using a directional laser system. The primary goal of the PaddleSat is to supplement power on a host spacecraft to reduce the requirements for onboard power systems of the host spacecraft or extend mission lifetimes. System performance estimates, link budget calculations, optical transmission hardware and link analysis, design tradeoffs between beam divergence, optical efficiency, and relative orbital control requirements are examined.
\end{abstract}

% Note that keywords are not normally used for peerreview papers.

\section{Introduction}

\IEEEPARstart{S}{atellites} operating in Low Earth Orbit (LEO) experience periodic eclipses during each orbit, which interrupt solar energy collection and place heavy cycling demands on onboard batteries. Over time, this repeated charge-discharge cycling degrades battery performance and reduces mission lifetime. To mitigate this issue, this project proposes the concept of an optical "charging companion" spacecraft, a PaddleSat, that can wirelessly transmit optical power to a host satellite during eclipse periods. 

The PaddleSat system would employ a directed laser or optical beam to deliver energy to the host's photovoltaic receiver, supplementing its power system when sunlight is unavailable. This concept, if demonstrated feasible, could extend mission lifetimes for small satellites, enable more continuous payload operations, and reduce the required battery capacity for future missions. 

The objectives of this project are to evaluate the feasibility of optical power transfer between two LEO spacecrafts, perform a detailed optical and RF link analysis, and propose a conceptual spacecraft design that can support this function. The expected outcomes include system performance estimates, link budget calculations, optical transmission hardware and link analysis, design tradeoffs between beam divergence, optical efficiency, and relative orbital control requirements. 

\section{Spacecraft Overview}\label{sec:spacecraft_overview}
The PaddleSat system is a lightweight, thin-film companion spacecraft designed to operate in close formation with a designated host satellite in Low Earth Orbit (LEO). Its primary function is to serve as an optical charging station, transmitting directed optical energy toward the host during eclipse periods to supplement its onboard power system. Unlike traditional rigid spacecraft, the PaddleSat utilizes a deployable membrane structure—similar to a solar sail or thin-film solar array—to maximize optical surface area while minimizing overall mass and structural complexity.\cite{ADSSFSS2017}
\subsection{PaddleSat Spacecraft Design and Environment}
The PaddleSat is a CubeSat class-nano spacecraft that provides electrical power, pointing and stability, communication, and station keeping required to support laser-based power delivery to the host satellite in LEO. A 6U configuration (10-11 kg) wet mass is selected to accommodate the optical transmission fine-pointing hardware, deployable power system, and required avionics. The spacecraft incorporates a 1 m$^2$ deployable photovoltaic array capable of producing approximately 340 W of peak power under direct solar illumination. Accounting for the eclipse periods and nominal bus loads, this corresponds to an average expendable power budget of 175-210 W, sufficient to operate a high-duty cycle laser transmitter with an electrical input of ~175 W. Energy storage is provided by a ~278 Wh-Li-ion battery, dimensioned to sustain continuous operation during eclipse periods and peak transmission intervals. The PaddleSat maintains a relative separation of 100-500 meters from its host satellite. \cite{ESA/PROBA-3}

Precision pointing and attitude control are achieved through an ADCS suite comprising of reaction wheels, magnetorquers, a star tracker, and an inertial measurement unit, augmented by a fast-steering mirror capable of maintaining residual pointing errors \(\le 20~\mu\mathrm{rad}\) over the nominal 100-500 meter separation. The spacecraft is further equipped with a micro-propulsion system for station-keeping maneuvers, with a $\Delta V$ budget on the order of 1--10~m/s and a S-band cross-link to enable command, telemetry, and relative navigation. Thermal management features are integrated to dissipate heat from the laser driver and other high-power subsystems. Collectively, these design elements enable the PaddleSat to maintain formation flight with the host satellite while reliably delivering optical power with meeting the pointing requirements. 

\subsection{Paddle Design}
Each paddle functions as the primary optical interface, converting stored electrical energy into a directed laser beam for power transfer. The paddle is a lightweight, thin-film membrane that supports a laser diode array and beam-forming optics. The transmitting aperture (10-15 cm diameter) is designed to minimize diffraction and beam divergence, and the laser operates at 980 nm to match the host satellite's solar cell efficiency. 

Beam pointing combines coarse spacecraft attitude control with a fine-steering mirror on each paddle, compensating for residual jitter to maintain alignment within \(20-100~\mu\mathrm{rad}\). The membrane is fabricated from space-qualified low-areal-density materials to ensure stability under thermal cycling and vibration while keeping each paddle below ~1 kg. Passive thermal features, such as high-emissivity coatings, dissipate heat from the laser diodes. Together, the two paddles provide a compact, high-efficiency optical interface capable of reliably delivering continuous-wave laser power over 100-500 meter distances, meeting the operational objectives of the PaddleSat mission. 

\begin{figure}[!h]
  \centering
  \includegraphics[width=0.9\columnwidth]{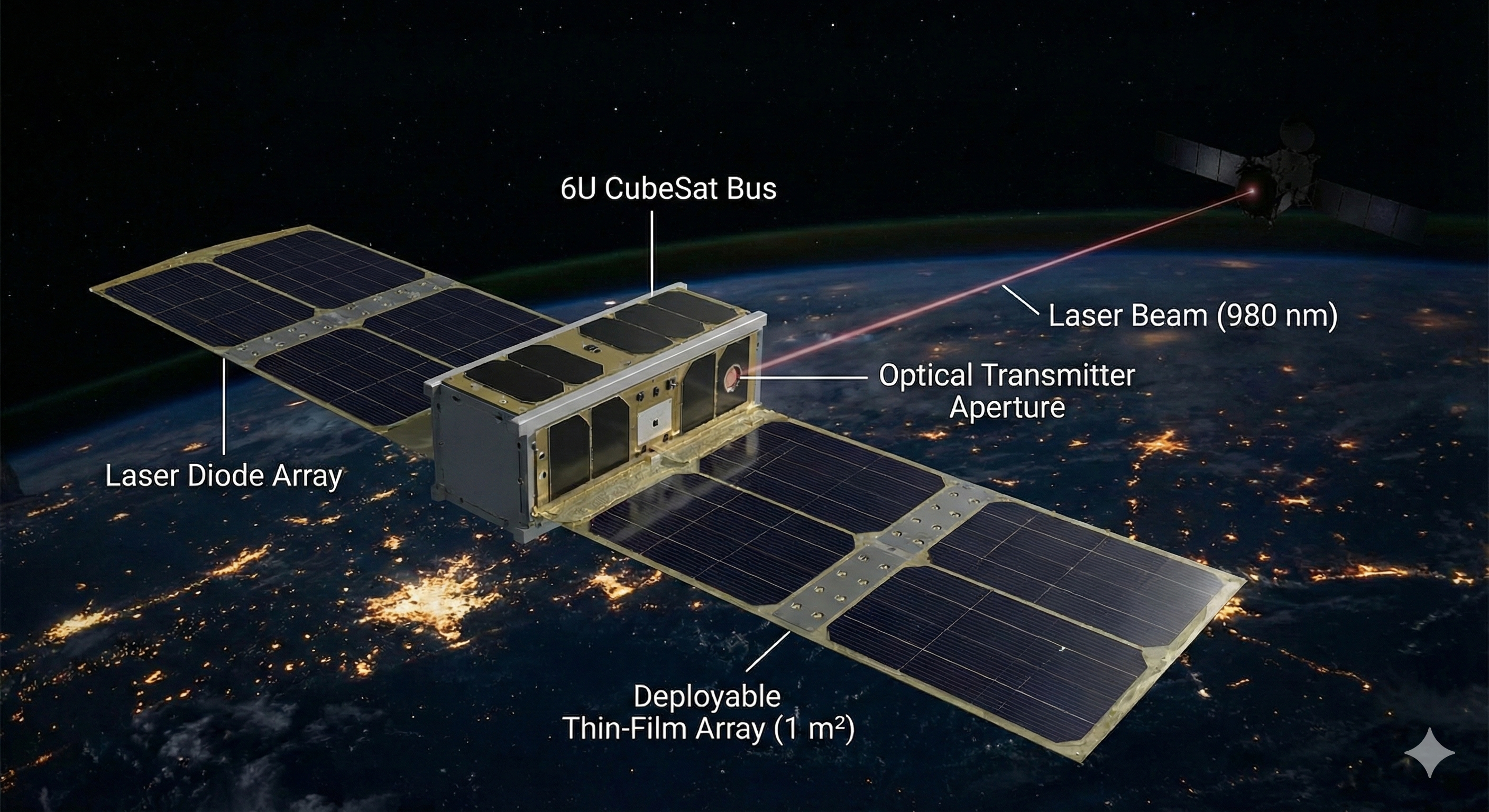}
  \caption{Conceptual rendering of the PaddleSat optical charging companion: a 6U CubeSat-class spacecraft with a 1 m$^2$ deployable thin-film photovoltaic array and laser diode array transmitting a 980 nm beam to a nearby host satellite at a nominal 100--500 m separation (not to scale).}
  \label{fig:paddlesat_concept}
\end{figure}

\subsection{Orbit Design and Station-Keeping}

The PaddleSat is deployed into a near-circular Low Earth Orbit matched to the host spacecraft to minimize differential perturbations and simplify formation maintenance. A nominal altitude of 500~km is selected, providing an orbital period of approximately 95~minutes and a mean motion of 
\[
n \;=\; \sqrt{\frac{\mu}{a^3}} \approx 1.1065\times 10^{-3}\ \mathrm{rad/s},
\]
for a circular orbit with $a = 6878$~km. Maintaining $e \approx 0$ reduces periodic variations in relative geometry and keeps the rotating Hill reference frame sufficiently stable for optical power transfer operations at ranges of 100--500~m.

Relative motion between PaddleSat and the host is described using the Hill--Clohessy--Wiltshire (HCW) equations.\cite{vallado2013} A key implication is the sensitivity of along-track drift to small semimajor-axis errors. A semimajor-axis bias $\Delta a$ produces a mean-motion offset 
\[
\Delta n \;\approx\; -\frac{3}{2}\,n\left(\frac{\Delta a}{a}\right),
\]
resulting in an along-track drift speed
\[
v_{\mathrm{drift}} \;\approx\; \frac{3}{2}\,n\,\Delta a.
\]
For $a = 6878$~km, this becomes $v_{\mathrm{drift}} \approx 1.66\times 10^{-3}\,\Delta a$ (m/s). Thus, a 1~m bias leads to $\sim$143~m/day of drift, whereas a 10~m bias produces $\sim$1.43~km/day. These values indicate that meter-scale orbit errors can cause kilometer-scale separation within hours to days, underscoring the need for regular station-keeping to maintain the 100--500~m operating envelope.

Station-keeping maneuvers counter this natural drift. Using the $\Delta a = 10$~m example, a daily correction of $\sim 0.0166$~m/s corresponds to an annual $\Delta V \approx 6.1$~m/s. Accounting for differential drag, $J_2$ perturbations, navigation errors, and collision-avoidance margins, a practical annual station-keeping requirement of 1--10~m/s is consistent with prior LEO formation-flying missions. \cite{VISORS2021} Propellant mass requirements follow the rocket equation:
\[
m_p = m_0\left(1 - e^{-\Delta V/(g_0 I_{sp})}\right).
\]
For a 6U spacecraft with $m_0 = 11$~kg and $\Delta V = 5$~m/s per year:
\begin{itemize}
    \item Cold gas ($I_{sp} = 60$~s): $m_p \approx 0.093$~kg/year,
    \item Electric micropropulsion ($I_{sp} = 1000$~s): \\ 
    $m_p \approx 0.0056$~kg/year.
\end{itemize}
These calculations show that both cold-gas and electric propulsion systems are feasible, with high-$I_{sp}$ systems offering significantly lower propellant mass at the expense of increased power and subsystem complexity.
Relative-navigation performance drives safety and control requirements. GNSS and S-band two-way ranging provide coarse relative-state estimates at the 1--10~m level, while optical beacons and onboard vision processing improve this to sub-meter accuracy during operations. Attitude control and fine beam pointing are maintained using reaction wheels, a star tracker, and the PaddleSat's fast-steering mirror. A pointing residual of 20~$\mu$rad corresponds to a linear displacement of
\[
\Delta r = \delta\theta\,L,
\]
giving 2~mm at $L = 100$~m and 10~mm at $L = 500$~m, which is small relative to photovoltaic receiver dimensions.

Operational safety is maintained through staged proximity operations: coarse phasing via GNSS/S-band, optical acquisition at kilometer scales, and controlled closure to the 100--500~m operating zone using closed-loop optical navigation. Laser transmission is permitted only when fault-protected arm/inhibit conditions are met, including verified relative-state bounds and confirmed pointing lock. An emergency retreat $\Delta V$ of 0.5--1.0~m/s provides rapid collision avoidance capability, requiring only $\mathcal{O}(10\ \mathrm{g})$ of cold-gas propellant. Collectively, these orbit design parameters and station-keeping calculations demonstrate that the PaddleSat can reliably maintain formation at 100--500~m separation while supporting the tight pointing and safety constraints required for optical power transfer.

\section{Optical Transmission Hardware}
The Optical Transmission Hardware is the primary system responsible for converting the PaddleSat's stored electrical power into a directed beam of energy. The design of this system is a trade-off between efficiency, mass, and pointing accuracy. The key components include the laser source and the beam-forming optics (transmitting telescope). 
\subsection{Laser Frequency (Wavelength) and Pulse Rate}
The primary objective is to select a laser that maximizes the power conversion efficiency of the host satellite's existing solar cells. Standard silicon (Si) photovoltaic cells, common in many satellite designs, have their peak quantum efficiency in the near-infrared (NIR) spectrum, typically between 850 nm and 1000 nm. Higher-efficiency cells, like Gallium Arsenide (GaAs), operate optimally at slightly shorter wavelengths, around 850-900 nm \cite{PTE1}. High-efficiency, space-qualified laser diodes are readily available in this 850-980 nm range. We propose a baseline wavelength of 980 nm due to the wide availability of high-power, high-efficiency fiber-coupled diodes at this wavelength, which also aligns well with the response curve of Si cells. Although 1550 nm sources offer improved beam safety and atmospheric transparency, 980 nm maximizes photovoltaic conversion for both Si and GaAs receivers while retaining mature, efficient, space-qualified diode technology. The final selection must be optimized for the specific photovoltaic material on the host satellite.

For raw power transfer, a Continuous Wave (CW) laser is proposed. A CW laser provides a constant, uninterrupted stream of energy, which is ideal for "charging" a solar cell as it mimics a steady illumination source. Pulsed lasers, while useful for data transmission or applications requiring high peak power (like LIDAR), are unnecessarily complex for this mission. A CW system simplifies the transmitter design and, more importantly, requires no specialized receiving electronics on the host satellite, as the existing solar panel is the receiver.

\subsection{Mirror / Lens Size and Weight}
The size (aperture) of the primary transmitting mirror or lens is the most critical factor determining the laser's dispersion (beam divergence). A larger aperture results in a more collimated beam, concentrating more power on the target. Given the "ultra-lightweight" requirement, a large, heavy mirror (like on a telescope) is not feasible. However, a larger aperture is essential for minimizing beam spread over the 100-500 m operational distance. We propose a baseline transmitter aperture diameter of 10-15 cm. 

The system mass will be minimized by fabricating the mirror from lightweight, space-qualified materials. A lightweight SiC-composite or aluminized polyimide membrane mirror is preferred; both have demonstrated sub-micron figure stability in microgravity at areal densities under 10 g/m² (e.g., JAXA IKAROS solar sail heritage). A 0.15m SiC composite mirror  has a mass around 0.12 kg; with mounts and actuators, the total optical assembly mass should be under 0.4 kg.

\subsection{Beam Divergence}
Beam divergence defines how much the laser beam spreads as it travels from the PaddleSat to the host. This, combined with pointing accuracy, determines the "spillover" loss—the amount of power that misses the target.

The theoretical, diffraction-limited divergence (the minimum possible spread) is inversely proportional to the aperture diameter. For a 15~cm aperture ($D = 0.15 \text{ m}$) and a 980~nm wavelength ($\lambda = 9.8 \times 10^{-7} \text{ m}$), the ideal divergence ($\Theta$) is:
\begin{equation}
\Theta \approx 2.44 \frac{\lambda}{D}
       \approx 2.44 \frac{9.8\times10^{-7}}{0.15}
       \approx 1.6\times10^{-5}\,\text{rad} \approx 16~\mu\text{rad}.
\end{equation}

This ideal value would result in a spot size of only $\sim$0.8 cm at 500 m. However, real-world systems are limited by imperfections (beam quality $M^2 > 1$) and, most significantly, the pointing jitter of the spacecraft.
Therefore, the design must accommodate a more realistic divergence to ease pointing requirements. We baseline a total system divergence (including jitter) of \textbf{100--200 \textmu rad}. At the maximum 500 m range, this divergence yields a spot diameter:
\begin{equation}
    \text{Spot size at } 500\,\text{m} 
    \approx 200~\mu\text{rad} \times 500\,\text{m} 
    = 0.10\,\text{m} \; (10\,\text{cm}).
\end{equation}
This 10~cm spot size is an excellent compromise. It is small enough to be captured efficiently by a single solar panel segment but large enough to be a feasible target for a realistic, low-cost pointing and tracking system.

\section{Optical Link Analysis}
The optical link, together with the transmission hardware, determines the delivered
power at the host photovoltaic receiver and therefore overall system effectiveness.
In LEO, atmospheric attenuation is negligible, so link performance is primarily
driven by (i) geometric beam spreading, (ii) pointing and jitter, and (iii) Doppler
shift \cite{POLB, L-L_OISL}.

\subsection{Geometric spreading (effective FSPL)}
At optical frequencies, propagation loss is dominated by beam divergence rather than
RF-style free-space path loss. For a diffraction-limited circular aperture, the
minimum divergence is
\begin{equation}
\Theta \approx 2.44\frac{\lambda}{D},
\end{equation}
but in practice the effective divergence is increased by non-ideal beam quality and
spacecraft pointing stability. The resulting spot size therefore grows approximately
linearly with range.

\subsection{Pointing error and jitter}
Pointing stability typically dominates the capture efficiency at 100--500~m
separation. Let $\sigma_{\theta,\mathrm{tx}}$ and $\sigma_{\theta,\mathrm{rx}}$ denote the
(one-sigma) angular jitter contributions from the transmitter and receiver spacecraft,
respectively. Assuming independent jitter sources, the net angular error is
\begin{equation}
\sigma_{\theta} \approx \sqrt{\sigma_{\theta,\mathrm{tx}}^2+\sigma_{\theta,\mathrm{rx}}^2}.
\end{equation}
This corresponds to a lateral displacement on the receiver plane of
\begin{equation}
\sigma_r \approx \sigma_{\theta}\,L,
\end{equation}
where $L$ is the inter-spacecraft separation. As a conservative illustration, using
$\sigma_{\theta,\mathrm{tx}}=250~\mu\mathrm{rad}$ and $\sigma_{\theta,\mathrm{rx}}=1000~\mu\mathrm{rad}$
yields $\sigma_r \approx 0.515$~m at $L=500$~m. In this regime, receiver array size and
closed-loop fine pointing (e.g., fast steering mirror) drive the fraction of power
captured, and therefore dominate the link efficiency \cite{nasa2021jitter}.

\subsection{Doppler shift}
Relative motion between two LEO spacecraft can introduce Doppler shift in the optical
carrier. However, the corresponding wavelength shift is typically small compared with
the spectral response width of silicon solar cells, and is generally less constraining
than pointing stability for power-transfer operation \cite{KHALID2024111033, PV_RES}.

Overall, assuming the receiver aperture/array is sized to accommodate the expected
spot growth and pointing dispersion, optical propagation losses are driven mainly by
pointing/jitter and receiver geometry rather than atmospheric effects.

\section{RF Telemetry Link Analysis}
The RF Command \& Telemetry (C\&T) link provides the control and observability needed
to safely operate the optical charging concept described in Sections~3--5. During eclipse,
the PaddleSat must (i) confirm that it is in the correct relative geometry from 
Section~\ref{sec:spacecraft_overview}, (ii) receive authorization to arm or inhibit
the optical beam, and (iii) report health and basic navigation state back to the host.
All of this traffic flows over a short-range S--band link between the PaddleSat and the
host satellite. 

In this work we only size the intra-orbit crosslink. The downlink to Earth can
follow standard CubeSat practice \cite{Liu_CubeSat_Antennas} and is not expected to be limiting. At the 
100--500~m separations considered in the formation design, the RF link behaves as a
simple free-space channel with negligible atmospheric or ionospheric loss in S--band. \cite{ESA_SBand_Freq}

\subsection{Crosslink Architecture and Assumptions}
The RF crosslink is intentionally simple, consistent with small-satellite and formation
flying architectures in \cite{CCSDS401,PrattBostian,LEOSatFormforSSP,Balanis}.  
Key assumptions are:
\begin{itemize}
    \item \textbf{Geometry:} relative separation 
          $R \in [100, 500]~\mathrm{m}$ as defined in the formation analysis.
    \item \textbf{Frequency band:} S--band, centered near $f = 2.3~\mathrm{GHz}$.
    \item \textbf{Data rate:} low-rate telemetry/command link, 
          $R_b = 19.2~\mathrm{kbps}$ BPSK/QPSK.
    \item \textbf{Traffic pattern:} short command and status messages, 
          duty-cycled around optical charging operations.
    \item \textbf{Receiver:} typical small-satellite RF front-end with 
          noise figure $\mathrm{NF} \approx 3~\mathrm{dB}$.
\end{itemize}

We use a half-duplex TDMA scheme where PaddleSat and host transmit in assigned time
slots. This avoids self-interference on a compact bus and matches the low average
throughput needs of this mission. \cite{proakis-digcomm}\cite{pratt-satcom}

Physically, both spacecraft carry compact S--band antennas mounted to preserve 
line-of-sight through the nominal relative orientations defined in 
Section~\ref{sec:spacecraft_overview}. At the very short ranges involved, the link
must work even if both antennas are near-isotropic.

\subsection{Free--Space Path Loss at Formation Separation}
For a space-to-space S--band link, the free--space path loss (FSPL) in dB is
\begin{equation}
    \mathrm{FSPL(dB)} = 
      92.45 
      + 20\log_{10}\!\left(f_{\mathrm{GHz}}\right)
      + 20\log_{10}\!\left(R_{\mathrm{km}}\right),
\end{equation}
where $f_{\mathrm{GHz}}$ is in GHz and $R_{\mathrm{km}}$ is the separation in km.

At the worst-case separation $R_{\max} = 500~\mathrm{m} = 0.5~\mathrm{km}$ and
$f = 2.3~\mathrm{GHz}$,
\begin{align}
    \mathrm{FSPL}_{\max}
      &\approx 92.45 + 20\log_{10}(2.3) + 20\log_{10}(0.5) \\
      \nonumber &\approx 93.7~\mathrm{dB}.
\end{align}

At these distances, atmospheric loss, scintillation, and rain fading are negligible
compared to $\sim 94~\mathrm{dB}$ of geometric path loss, so we ignore them.

\subsection{Sizing Transmit Power and Antenna Gain}
We now back-solve for the transmit power and antenna gains required to close the link
with comfortable margin, following the style of small-satellite link budgets in
\cite{CCSDS401,PrattBostian,LEOSatFormforSSP}.

\subsubsection{Required $E_b/N_0$ and Received Power}
For coded BPSK/QPSK telemetry links at bit error rates on the order of 
$\mathrm{BER} \le 10^{-5}$, a required energy-per-bit to noise-density ratio of
\begin{equation}
    \left(\frac{E_b}{N_0}\right)_{\mathrm{req}} \approx 9.6~\mathrm{dB}
\end{equation}
is typical when using CCSDS-compliant FEC \cite{CCSDS401}. We also allocate an
explicit link margin of
\begin{equation}
    M_{\mathrm{link}} = 10~\mathrm{dB},
\end{equation}
consistent with conservative formation-flying studies in LEO.

The thermal noise spectral density at the receiver input is
\begin{equation}
    N_0(\mathrm{dBm/Hz}) \approx -174~\mathrm{dBm/Hz} + \mathrm{NF(dB)},
\end{equation}
so with $\mathrm{NF} = 3~\mathrm{dB}$ we use $N_0 \approx -171~\mathrm{dBm/Hz}$.

The carrier power at the receiver required to meet the $E_b/N_0$ target plus margin is
\begin{align}
    C_{\mathrm{req}}(\mathrm{dBm})
      = & \left(\frac{E_b}{N_0}\right)_{\mathrm{req}} 
      + M_{\mathrm{link}}
      + 10\log_{10}\!\big(R_b\big) \\
      \nonumber & + N_0(\mathrm{dBm/Hz}),
\end{align}
where $R_b$ is in bits/s. With $R_b = 19.2~\mathrm{kbps}$,
\begin{align}
    C_{\mathrm{req}} 
      &\approx 9.6 + 10 + 10\log_{10}(1.92\times 10^4) - 171 \\
      \nonumber &\approx -109~\mathrm{dBm}.
\end{align}

Any received carrier above about $-109~\mathrm{dBm}$ satisfies both the error-rate
requirement and our built-in link margin.

\subsubsection{Back-Solved EIRP}
Let $G_{\mathrm{tx}}$ and $G_{\mathrm{rx}}$ be the transmit and receive antenna gains, and
let $L_{\mathrm{sys}}$ collect cable, mismatch, and small pointing/polarization losses.
We take $L_{\mathrm{sys}} \approx 3~\mathrm{dB}$. The received power is
\begin{equation}
    P_r = P_{\mathrm{tx}} + G_{\mathrm{tx}} + G_{\mathrm{rx}} - \mathrm{FSPL} - L_{\mathrm{sys}},
\end{equation}
Setting $P_r = C_{\mathrm{req}}$ and solving for required transmit EIRP gives
\begin{equation}
    \mathrm{EIRP}_{\mathrm{req}} 
      = P_{\mathrm{tx}} + G_{\mathrm{tx}}
      = C_{\mathrm{req}} + \mathrm{FSPL}_{\max} + L_{\mathrm{sys}} - G_{\mathrm{rx}}.
\end{equation}

For a conservative case with effectively isotropic antennas on both spacecraft
($G_{\mathrm{tx}} \approx G_{\mathrm{rx}} \approx 0~\mathrm{dBi}$),
\begin{align}
    \mathrm{EIRP}_{\mathrm{req}}
      &\approx -109~\mathrm{dBm} + 93.7~\mathrm{dB} + 3~\mathrm{dB} - 0 \\
      \nonumber &\approx -12~\mathrm{dBm}.
\end{align}
An EIRP of $-12~\mathrm{dBm}$ corresponds to roughly $60~\mathrm{\mu W}$ of radiated
power. This shows that at 500~m separation, the link is extremely forgiving compared
to typical inter-satellite links at hundreds or thousands of kilometers 
\cite{LEOSatFormforSSP}.

\subsubsection{Baseline Hardware Choice and Margin}
Rather than designing to the absolute minimum, we choose a simple, robust configuration
that strongly exceeds the requirement:
\begin{itemize}
    \item PaddleSat S--band transmitter: $P_{\mathrm{tx}} = 0~\mathrm{dBm}$ (1~mW).
    \item PaddleSat antenna: small RHCP patch, $G_{\mathrm{tx}} \approx 3~\mathrm{dBi}$.
    \item Host antenna: low-gain omni or mini-patch, $G_{\mathrm{rx}} \approx 0~\mathrm{dBi}$.
\end{itemize}
At $R_{\max}$ this yields
\begin{align}
    P_r 
      &\approx 0 + 3 + 0 - 93.7 - 3 \\
      \nonumber &\approx -93.7~\mathrm{dBm}.
\end{align}

For a receiver bandwidth of $B = 100~\mathrm{kHz}$, the noise power is
\begin{align}
    N(\mathrm{dBm})
      &= -174 + 10\log_{10}(B_{\mathrm{Hz}}) + \mathrm{NF} \\
      \nonumber &\approx -174 + 50 + 3 \\
      \nonumber &\approx -121~\mathrm{dBm},
\end{align}
so the carrier-to-noise ratio is
\begin{equation}
    \frac{C}{N}(\mathrm{dB}) \approx P_r - N \approx -93.7 - (-121) \approx 27~\mathrm{dB}.
\end{equation}
The corresponding $E_b/N_0$ is
\begin{align}
    \frac{E_b}{N_0}(\mathrm{dB})
      &= \frac{C}{N}(\mathrm{dB}) 
       + 10\log_{10}\!\left(\frac{B}{R_b}\right) \\
      \nonumber &\approx 27 + 10\log_{10}\!\left(\frac{100~\mathrm{kHz}}{19.2~\mathrm{kbps}}\right) \\
      \nonumber &\approx 34~\mathrm{dB},
\end{align}
which exceeds the required $\sim 9.6~\mathrm{dB}$ by more than $20~\mathrm{dB}$ even after
our explicit $10~\mathrm{dB}$ link margin.

In other words, a 1~mW S--band transmitter with a very small patch antenna is already
vastly overdesigned for the 100--500~m PaddleSat--host geometry. In the context of
this mission, the RF telemetry load on the power system is negligible compared to the
optical transmitter and the rest of the spacecraft bus.

\subsection{Integration with Optical Charging Operations}
Because the crosslink closes with tens of dB of margin using milliwatt-level transmit
power and near-isotropic antennas, the RF subsystem does not constrain:
\begin{itemize}
    \item the allowable PaddleSat--host separation range from the formation analysis,
    \item the optical aperture pointing trades in Section~3, or
    \item the power budget and storage sizing in Sections~2 and~5.
\end{itemize}
Instead, the RF system can be implemented with COTS-class radios and very simple
antennas, while still providing reliable arm/inhibit control and status feedback during
each optical charging session. This allows the design effort to focus on the optical
link, formation control, and thin-film structures that dominate the overall mission
performance.

\section{Power Transfer Efficiency}
This section overviews details of the optical laser's power transfer efficiency, as well as potential losses that factor into decreased efficiency. Power Transfer Efficiency (PTE) refers to the percentage of received power with respect to transmit power. Total power transfer efficiency also equals the product of efficiencies of the laser emitter, laser transmission, and the laser receivers. Overall PTE estimates are used with the proposed photovoltaic cell array and battery selection in order to calculate power provided to the host satellite.

The greatest factor in calculating power transfer efficiency is placed on the laser emitter, which involves selection of hardware to be installed on the PaddleSat. The emitter must meet low size, weight, and power (SWaP) considerations of the PaddleSat, while also maintaining a lifetime that meets mission needs. Within the last 30 years, lasers operating at the selected 980 nm wavelength have a nominal PTE around 25\%, with a highest recorded PTE of 51\% \cite{PTE3}. The selected laser emitter to meet highest PTE is a VCSEL, or vertical-cavity surface-emitting laser. VCSELs use two semiconductor Bragg mirrors with 99.9\% reflectivity (no efficiency loss here). They are notable for having low beam divergence, which enables them to be used at range in this application. An significant benefit to VCSEL selection is the ability to easily scale VCSELs in an array. VCSEL arrays can be spaced within tens of microns from each other, thus increasing the overall output power of the emitter while simultaneously increasing redundancy and lifespan on the PaddleSat (if one VCSEL fails, it's only a small percentage of the overall emitter array) \cite{PTE5}. VCSEL arrays have already been proposed and tested in CubeSats, with an example 19-element 980 nm VCSEL array fabricated by NASA, which yielded an output power of 150 mW with 30\% efficiency \cite{PTE6}.

After the emitter generates the laser, the next component in overall power transfer efficiency is the laser transmission. There are a few notable losses that may affect laser transmission and thus lower PTE. First is reflectivity, which is negligible in VCSEL arrays due to their near ideal Bragg mirrors. Other major bottlenecks of laser transmission are less concerning in LEO, making laser transmission in space a relatively efficient selection. Atmospheric effects, including absorption, scattering, turbulence, and nonlinear effects, all of which impact ground-based laser transmission, are negligible in a LEO application \cite{PTE2}. Despite satellites in the thermosphere (LEO altitude) being subject to atmospheric drag, the density equates to nearly zero resulting in the assumption of no atmospheric path loss \cite{PTE7}. The largest potential loss comes from aiming accuracy error, which depends on beam divergence, maneuverability of the PaddleSat, and spot diameter of the receiver.

The last factor in power transfer efficiency is the receiver. The selected receiver is a silicon (Si) photovoltaic cell. Overall power transfer efficiency for Silicon photovoltaic cells trends around 20-25\%, and even lower in space due to decreased air mass (one source noting 14.8\% in 2008) \cite{PTE7}. However, when applying monochromatic illumination during the eclipsing period of a host satellite, this loss of air mass is irrelevant. Considering monochromatic illumination, Si cells are highly recommended. At 980 nm wavelength, the external quantum efficiency of an Si cell is over 92\% and results in an efficiency near 35\% \cite{PTE4}.

\begin{table}[h]
\centering
\caption{PTE Percentages}
\label{pte-percentages}
\begin{tabular}{|c|c|}
\hline
\textbf{Factor} & \textbf{\% Efficiency} \\ \hline
Emitter         & 30\%                   \\ \hline
Transmission    & 99.9\%                 \\ \hline
Receiver        & 35\%                   \\ \hline
\end{tabular}
\end{table}

Based on the above analysis, the percentages in Table \ref{pte-percentages} are used for calculation in overall power transfer efficiency. 

\begin{equation}
\eta_{total} = \eta_{emitter} * \eta_{transmission} * \eta_{receiver}
\end{equation}

\begin{equation}
\eta_{total} = 0.3 * 0.999 * 0.35 = 0.105
\end{equation}

The overall power transfer efficiency of the system is thus 10.5\%. 

\subsection{Transmit Power \& Battery Capacity}
Total transmit power of the PaddleSat depends on its solar panel and battery capacity. We propose a 1\(m^2\) solar panel array using a single crystalline Silicon photovoltaic cell. The photovoltaic cell properties are defined in Table \ref{cell-properties}. An optimal incident angle has been selected for these calculations in order to focus efforts on hardware selection rather than maneuverability of the PaddleSat.

\begin{table}[h]
\centering
\caption{Photovoltaic Cell Properties}
\label{cell-properties}
\begin{tabular}{|c|c|}
\hline
Panel Area       & 1\(m^2\)   \\ \hline
Panel Efficiency & 25\% \\ \hline
Solar Irradiance & 1362\(\frac{W}{m^2}\) \\ \hline
Incident Angle & 0\(^\circ\) \\ \hline
\end{tabular}
\end{table}

The photovoltaic cell properties are used to calculate the amount of power received by the PaddleSat's solar panel

\begin{equation}
    P = A\times G_{sc}\times \eta \times cos\theta
\end{equation}

where \(A\) is the solar panel's area, \(G_{sc}\) is the solar irradiance constant, \(\eta\) is the panel's conversion efficiency, and \(\theta\) is the incident angle. Using the Si cell properties above, the power received by the PaddleSat's solar panel is \(340.5W\).  

Assuming a LEO altitude of \(500km\) with a circular orbit, the orbital period equals

\begin{equation}
    T = \sqrt{\frac{4\pi^2R^3}{GM_e}}
\end{equation}

where \(R\) is the radius of the PaddleSat's orbit, \(G\) is the gravitational constant, and \(M_e\) is the Earth's mass. The orbital period is approximately 5679 seconds (just under 95 minutes). Correspondingly, the eclipse period equals

\begin{equation}
    T_e = \frac{T}{2} - \frac{Tcos^{-1}(\frac{R_e}{R})}{\pi}
\end{equation}

where \(R_e\) is the Earth's radius. The eclipse period of the PaddleSat is approximately 2146 seconds (35 minutes, 46 seconds).

The total average power to be expended by the PaddleSat (including charging its battery) equals the amount of power that solar panel receives during the daylight period. 

\begin{equation}
\begin{split}
    P_{expend} & = P \times \frac{T-Te}{T} = 340.5 \times \frac{5679-2146}{5679} \\
    & = 212 W
\end{split}
\end{equation}

Thus, the PaddleSat can expend on average 212\(W\) of power throughout its orbit, which equals 334\(Wh\) of energy. This energy expenditure accounts for charging a battery that has enough capacity to equal these power demands, but does not yet account for any spare capacity of the battery. Thus, battery capacity will be calculated to match the energy expenditure, and overall energy expenditure should be updated to account for additional spare capacity of the battery.

The battery selected for the PaddleSat is a Lithium-Ion \((LiCoO_2)\) battery with specifications defined in Table \ref{li-properties}.

\begin{table}[h]
\centering
\caption{Lithium-Ion Battery Properties}
\label{li-properties}
\begin{tabular}{|c|c|}
\hline
Energy Density    & 200\(\frac{W-hr}{kg}\)  \\ \hline
Potential Voltage & 4\(V\)   \\ \hline
Spare Capacity    & 20\% \\ \hline
\end{tabular}
\end{table}

The spare capacity of this battery is a very conservative amount, given that Lithium Ion batteries can commonly handle greater than 95\% depth of discharge. 

\begin{equation}
\begin{split}
    M_{battery} & = \frac{E}{Energy Density} \\
    & = \frac{\frac{334 Wh}{1.2}}{200\frac{Wh}{kg}} \\
    & = 1.4 kg
\end{split}
\end{equation}

\begin{equation}
    \begin{split}
        Battery Capacity & = \frac{E}{PotentialVoltage} \\
        & = \frac{278Wh}{4V} \\
        & = 69.6 Ah
    \end{split}
\end{equation}

Therefore, accounting for the the 20\% spare battery capacity, the average power expenditure of the PaddleSat is 176\(W\) (corresponding to 278\(Wh\) of energy). This is made possible by including on the PaddleSat a Lithium Ion battery that weighs 1.4\(kg\) with a capacity of 69.6\(Ah\).

\subsection{Power Transfer to Host Satellite}
Given a budget of 176\(W\) of power, where less than 1\(W\) will be used for the RF telemetry link, this leaves 175\(W\) of power to be used for the optical charging link (assuming nominal operation not requiring power to other subsystems). With a calculated power transfer efficiency of 10.5\%, this corresponds to the host satellite receiving 18.375\(W\) of power from the PaddleSat's optical laser charging link.

\section{Conclusion}
 This project proposed and validated the concept of an optical "charging companion" spacecraft, a PaddleSat, that can wirelessly transmit optical power to a host satellite during eclipse periods using a laser. Optical power transfer between two LEO spacecraft has been shown to be feasible based off a detailed optical and RF link analysis, and a conceptual spacecraft design that can support this function. Spacecraft operating with the benefit of this system have been shown to have reduced demands on onboard batteries which may improve battery performance and increase mission lifetime.

\bibliographystyle{IEEEtran}
\bibliography{main}   

@techreport{CCSDS401,
  author       = {{Consultative Committee for Space Data Systems (CCSDS)}},
  title        = {Radio Frequency and Modulation Systems—Part 1: Earth Stations and Spacecraft},
  institution  = {CCSDS},
  number       = {401.0-B},
  type         = {Blue Book},
  year         = {2023},
  note         = {Latest revision},
}

@book{PrattBostian,
  author    = {Timothy Pratt and Charles W. Bostian and Jeremy E. Allnutt},
  title     = {Satellite Communications},
  edition   = {2},
  publisher = {Wiley},
  address   = {Hoboken, NJ},
  year      = {2003}
}

@techreport{PTE1,
  author       = {Geoffrey A. Landis},
  title        = {Photovoltaic Receivers for Laser Beamed Power in Space},
  institution  = {NASA},
  year         = {1991},
  month        = dec,
  note         = {NASA Technical Report},
  url          = {https://ntrs.nasa.gov/api/citations/19920006800/downloads/19920006800.pdf}
}

@book{Balanis,
  author    = {Constantine A. Balanis},
  title     = {Antenna Theory: Analysis and Design},
  edition   = {4},
  publisher = {Wiley},
  address   = {Hoboken, NJ},
  year      = {2016}
}

@techreport{LEOSatFormforSSP,
  author      = {Shu Ting Goh and Seyed Alireza Zekavat and Ossama Abdelkhalik},
  title       = {LEO Satellite Formation for SSP: Energy and Doppler Analysis},
  institution = {IEEE Transactions on Aerospace and Electronic Systems},
  year        = {2015},
  month       = jan,
  url         = {https://ieeexplore.ieee.org/stamp/stamp.jsp?tp=&arnumber=7073472}
}

@techreport{POLB,
  author      = {{Dhruv Shivkant and Shreyaans Jain and Rohit K Ramakrishnan}},
  title       = {Probabilistic Link Budget Analysis for Low Earth Orbit Satellites in the Optical Regime},
  institution = {Indian Institute of Science},
  year        = {2025},
  note         = {Latest revision},
  url = {https://doi.org/10.48550/arXiv.2507.20908}
}

@ARTICLE{L-L_OISL,
  author={Vieira, I. P. and Pita, T. C. and Mello, D. A. A.},
  journal={IEEE Access}, 
  title={Modulation and Signal Processing for LEO-LEO Optical Inter-Satellite Links}, 
  year={2023},
  volume={11},
  number={},
  pages={63598-63611},
  doi={10.1109/ACCESS.2023.3287501}}

@techreport{ESA/PROBA-3,
  author      = {Marta Casti and Silvano Fineschi and Vladimiro Noce and Cedric Thizy and Damien Galano},
  title       = {Fine Positioning Algorithims for the ESA/PROBA-3 formation flying mission},
  institution = {IEEE Transactions on Aerospace and Electronic Systems},
  year        = {2019},
  month       = jun,
  url         = {https://ieeexplore.ieee.org/document/8869551}
}

@techreport{PTE2,
  author = {Yifan Zheng},
  title = {Wireless laser power transmission: Recent progress and future challenges},
  institution = {ScienceDirect},
  year = {2024},
  month = jun,
  url = {https://www.sciencedirect.com/science/article/pii/S2950104023000020}
}

@inproceedings{PTE3,
    author = {Jean-Francois Seurin and Chuni L. Ghosh and Viktor Khalfin and Aleksandr Miglo and Guoyang Xu and James D. Wynn and Prachi Pradhan and L. Arthur D'Asaro},
    title = {{High-power high-efficiency 2D VCSEL arrays}},
    institution = {SPIE},
    volume = {6908},
    booktitle = {Vertical-Cavity Surface-Emitting Lasers XII},
    editor = {Chun Lei and James K. Guenter},
    organization = {International Society for Optics and Photonics},
    publisher = {SPIE},
    pages = {690808},
    year = {2008},
    doi = {10.1117/12.774126},
    URL = {https://doi.org/10.1117/12.774126}
}

@techreport{PTE4,
  author={Phillip Jenkins and Raymond Hoheisel and David Scheiman and Justin Lorentzen and Richard Fischer and David Wayne and  Brittany Lynn and Paul Jaffe},
  institution = {IEEE PVSC},
  booktitle={2018 IEEE 7th World Conference on Photovoltaic Energy Conversion (WCPEC) (A Joint Conference of 45th IEEE PVSC, 28th PVSEC & 34th EU PVSEC)}, 
  title={Silicon Solar Arrays for Laser Power Transfer Applications}, 
  year={2018},
  pages={3352-3356},
  doi={10.1109/PVSC.2018.8547719}
}

@online{PTE5,
  author       = {R. Paschotta},
  title        = {Vertical Cavity Surface-emitting Lasers},
  year         = {2006},
  organization = {RP Photonics AG},
  journaltitle = {RP Photonics Encyclopedia},
  url          = {https://www.rp-photonics.com/vertical_cavity_surface_emitting_lasers.html},
  urldate      = {2025-12-03},
  doi          = {10.61835/qcg}
}

@techreport{PTE6,
  author = {Peter M. Goorjian},
  title = {Space Optical Communications using Laser Arrays for CubeSats in Low Earth Orbit (LEO) or Low Lunar Orbit (LLO)},
  institution = {NASA Ames Research Center},
  year = {2020},
  month = oct,
  url = {\url{https://ntrs.nasa.gov/api/citations/20205008121/downloads/Goorjian\%20Tech\%20Transfer\%20webinar\%20Tu\%2010-13-2020\%20\%20-\%20\%20Read-Only.pdf}}

}

@techreport{PTE7,
    author = {Charlie T. Bellows},
    title = {Minimizing Losses in a Space Laser Power Beaming System},
    institution = {Air Force Institute of Technology},
    year = {2010},
    url = {https://apps.dtic.mil/sti/tr/pdf/ADA518829.pdf}
}

@misc{ESA_SBand_Freq,
  author       = {{European Space Agency}},
  title        = {Satellite Frequency Bands},
  year         = {2023},
  howpublished = {\url{https://www.esa.int/Applications/Connectivity_and_Secure_Communications/Satellite_frequency_bands}},
  note         = {Accessed: 2025-12-03}
}

@article{Liu_CubeSat_Antennas,
  author  = {Shuai Liu et al.},
  title   = {A Survey on CubeSat Missions and Their Antenna Designs},
  journal = {Electronics},
  volume  = {11},
  number  = {13},
  year    = {2022},
  note    = {Includes S-band (around 2.2~GHz) CubeSat downlink examples},
  url     = {https://www.mdpi.com/2079-9292/11/13/2021}
}

@techreport{nasa2021jitter,
  author       = {NASA},
  title        = {Spacecraft Line-of-Sight Jitter Management and Mitigation Lessons Learned and Engineering Best Practices},
  institution  = {NASA},
  number       = {NASA/TM-20210017871},
  year         = {2021}
}

@article{KHALID2024111033,
  title = {Characterization of Doppler shift in inter-satellite laser link between LEO, MEO, and GEO orbits},
  journal = {Optics \& Laser Technology},
  volume = {177},
  pages = {111033},
  year = {2024},
  issn = {0030-3992},
  doi = {https://doi.org/10.1016/j.optlastec.2024.111033},
  url = {\url{https://www.sciencedirect.com/science/article/pii/S0030399224004912}},
  author = {Muhammad Khalid and Wu Ji and Deng Li and Li Kun}
}

@techreport{PV_RES,
  title        = {Spectral Response and PV Modeling Guide},
author = {Sandia National Lab},
institution  = {Sandia National Lab},
  howpublished = {\url{https://pvpmc.sandia.gov/modeling-guide/2-dc-module-iv/effective-irradiance/spectral-response/}},
year = 2025,
  note         = {Accessed: 2025-12-04}
}

@book{vallado2013,
  title={Fundamentals of Astrodynamics and Applications},
  author={Vallado, David A.},
  year={2013},
  edition={4th},
  publisher={Microcosm Press}
}

@techreport{ADSSFSS2017,
  title= {Advanced Deployable Structural Systems For Small Satellites},
  author= {W.Keith Belvin and Marco Straubel and W. Keats Wilkie and Martin E. Zander and Juan M. Ferndandez and Martin F. Hillebrandt},
  institution= {NASA Langley Research Center},
  year= {2017},
  url= {\url{https://ntrs.nasa.gov/api/citations/20170003919/downloads/20170003919.pdf}}
}

@book{proakis-digcomm,
  author    = {Proakis, John G. and Salehi, Masoud},
  title     = {Digital Communications},
  edition   = {5th},
  year      = {2007},
  publisher = {McGraw-Hill}
}

@book{pratt-satcom,
  author    = {Pratt, Timothy and Bostian, Charles W. and Allnutt, Jeffrey E.},
  title     = {Satellite Communications},
  edition   = {2nd},
  year      = {2003},
  publisher = {Wiley}
}

@techreport{VISORS2021,
    author = {Adam W. Koenig and Simone D'Amico and E. Glenn Lightsey},
    title = {Formation Flying Orbit and Control Concept for VISORS Mission},
    institution = {Stanford University and Georgia Institute of Technology},
    year = {2021},
    url= {https://slab.sites.stanford.edu/sites/g/files/sbiybj25201/files/media/file/scitech_visors_2021_final.pdf}
}
\end{document}